\documentstyle[12pt,aasms4]{article}

\begin{document}

\title{BATSE OBSERVATIONS OF THE PICCINOTTI SAMPLE OF AGN}

\author{A.~Malizia \\
Department of Physics and Astronomy, University of Southampton, 
SO17  1BJ, England}

\author{L. Bassani\\
Istituto TeSRE/CNR, via Gobetti 101, 40129 Bologna, Italy}

\author{S.~N.~Zhang\\
University Space Research Association, Huntsville, AL 35806, USA}

\author{A.~J.~Dean\\
Department of Physics and Astronomy, University of Southampton, 
SO17 1BJ, England}

\author{W.~S.~Paciesas\\
University of Alabama, Huntsville, AL 35899, USA}

\author{G.~G.~C.~Palumbo\\
Universit\'a di Bologna, Astronomy Department Bologna, Italy}

\begin{abstract}

BATSE Earth occultation data have been used to search for emission in the 
20-100 keV band from all sources in the Piccinotti sample, which represents
to date the only complete 2-10 keV survey of the extragalactic sky down to a 
limiting flux of 3.1 $\times$ 10$^{-11}$ erg cm$^{-2}$ s$^{-1}$. 
Nearly four  years of
observations have been analyzed to reach a 5$\sigma$ sensitivity level
of about 7.8 $\times$ 10$^{-11}$ erg cm$^{-2}$ s$^{-1}$ in the band considered.
Of the 36 sources 
in the sample, 14 have been detected above 5$\sigma$ confidence level while
marginal detection (3$\le$ $\sigma$ $\le$5) can be claimed for 13 sources; 
for 9 objects 2 $\sigma$ upper limits are reported. 
Comparison of BATSE results with data at higher 
energies is  used to estimate the robustness of our data analysis:
while the detection level of each source is reliable, the flux measurement
maybe overestimated in some sources by as much as 35$\%$, probably due to
incomplete data cleaning. 
Comparison of BATSE fluxes with X-ray fluxes, 
obtained in the 2-10 keV range and averaged over years, indicates that a 
canonical power law of photon index 1.7 gives a good description 
of the broad band 
spectra of bright AGNs and that spectral breaks preferentially occur 
above 100 keV.
\end{abstract}


\section{Introduction}

The only statistically complete and large sample of X-ray sources in the 
energy range 2-10 keV was obtained using data from the A-2 experiment on 
the HEAO-1 satellite (Piccinotti et al. 1982). This instrument performed a 
survey of 8.3 sr of the sky (65.5\% coverage) at $|b| \ge20^{\circ}$. 
Sources have been included in the catalogue if detected with a 
significance greater than or equal to 5$\sigma$. The Piccinotti sample 
contains 85 sources (excluding the LMC and SMC sources) down to a limiting 
flux of 3.1 $\times$ 10$^{-11}$ erg
cm$^{-2}$ s$^{-1}$. Of the 68 objects of extragalactic origin, 36 have been 
identified with active galaxies while the remaining sources were found to 
be associated with clusters of galaxies. The AGN sample is composed of 30 
Seyfert galaxies (of which 23 are of type 1 and 7 of type 2), one starburst 
galaxy (M82), 4 BL Lac objects and one QSO (3C 273).
Some objects which were left unidentified in the original Piccinotti list 
are now identified (Miyaji and Boldt 1990, Giommi et al. 1991, Fairall,
McHardy and Pye 1982); these are 
H0111-149 (MKN1152, z=0.0536), H0235-52 
(ESO198-G24, z=0.045), H0557-385 (IRAS F05563-3820, z= 0.034), H0917-074 
(EXO0917.3-0722, z=0.169), H1829-591 (F49, z=0.02) and 
H1846-768 (1M1849-781, z=0.074). Only one object (H1325-020) is still 
unidentified but it is probably associated to a galaxy cluster 
(A1750, Marshall et al. 1979) and therefore has been excluded from the
present study. Three objects in the sample have the X-ray emission 
contaminated by a nearby source (IIIZW2 by the visual binary HD560, 
Tagliaferri et al. 1988, IRAS F05563-3820 by the BL Lac object EXO055625-
3838.6, Giommi et al. 1989 and 3C445 by the cluster of galaxies A2440, Pounds
1990); however, 
since these nearby objects should give a negligible contribution to the 
emission above 20 keV, none of these sources have been excluded
from the present analysis. 
\\
The Piccinotti sample is the hard X-ray selected sample
of AGN best studied at all wavelengths below 10-20 keV and as such has been 
used to study AGN X-ray spectral characteristics, LogN-logS relation 
and luminosity function and hence to 
determine the active galaxy contribution to the cosmic diffuse X-ray 
background in the 2-10 keV energy range.
In the X-ray band most of the AGNs in the sample have been observed
several times individually with different space borne experiments.
Spectral data in the 2-10 keV range obtained over the last 20 years has been
summarized in Ciliegi et al. (1993), Malaguti et al. (1994) and Malizia et al.
(1997) for all objects except H0557$-$385 and H0917$-$074; the 
spectrum of 
the former can be found in Turner et al. (1996) while no spectral data
have so far been reported for the latter object.
More recently, intrinsic absorption and 
soft excesses have been studied for 31 of the Piccinotti
sources (the 4 BL Lacs and the starburst galaxy M82 were not considered)
using all-sky survey data from the ROSAT
satellite in the 0.1-2 keV band (Schartel et al. 1997).
A systematic study of the X-ray variability of the sample 
has been performed by Turner \& Pounds (1989), Giommi et al. (1990) and
Grandi et al. (1992) 
with the Low Energy (LE) (0.05-2 keV) and Medium
Energy (ME) (2-50 keV) experiments onboard EXOSAT. Only in the last few years,
high energy observations of individual objects in the sample started to be
performed.
Prior to CGRO and BeppoSAX,
a handful of objects were detected above 10 keV by balloon
borne telescopes, HEAO A-4 and Sigma/GRANAT 
(Rothschild et al. 1983, Bassani et al. 1985, 
Bassani et al. 1993). 
More recently, data on most of these galaxies have  been
obtained both by OSSE/CGRO and BeppoSAX/PDS. 
27 of the Piccinotti objects has so far been observed by OSSE, 11 with 
detection above $\geq$ 5$\sigma$ level
(McNaron-Brown et al. 1995 and Johnson, 1997 private communication)
while BeppoSAX reported high energy emission from at least 16 sources in the
sample (Matt 1998, Bassani et al. 1998, Cappi et al. 1998, Pian et al. 1998,
Antonelli et al. 1998, BeppoSAX public archive).
The BATSE data reported here provide, for
the first time, a systematic coverage at high energies of the whole sample.
In this first work  BATSE data in the 20-100 keV range are reported for all 
sources (presentation of light curves and images are postponed to a future 
paper) and, whenever possible, comparisons with previous
soft and hard X-ray data have been performed.

\section{Observations and Results} 
The Burst and Transient Source Experiment (BATSE) onboard CGRO, is the first
all-sky monitor operating from 20 keV up to several MeV.
The BATSE daily monitoring sensitivity is $\sim$100 mCrab, implying that 
sensitivities of the order of a few mCrab could be obtained by integrating the 
data over a period of a few years, if systematic errors can be kept 
sufficiently small (McCollough et al. 1996).
The data used in this work were collected by the Large Area Detectors (LADs)
in Earth occultation mode (Harmon et al. 1992); this mode allows to measure a
source flux by looking at changes in the LAD background rates 
when the object under investigation rises from and sets behind the Earth.
A pair of rising and setting steps are generated
during each 90 minute satellite orbit implying that a large number of steps
must be summed to extract signal from weak sources such as extragalactic
objects. The Earth occultation method allows a nearly 
complete sky coverage (sources whose
declination is $\leq$35 degrees are always occulted, while for other
objects the coverage efficiency is 90\% or better) and so it is  
ideal to study moderately large samples of sources. Since the occultation 
technique reaches its peak sensitivity below
$\sim$ 140 keV, the search for emission by AGN in the Piccinotti
sample was carried out in the range between 20 and 100 keV. 
\\
BATSE data from nearly four years of observations (November 93 - September 97) 
have been analyzed to extract a signal from sources in the Piccinotti sample.
Interference from bright sources flaring or occulted at the same time
of the object being studied are the major source of 
systematic errors in the flux measurement and so careful cleaning of the 
data is necessary before estimating the  flux.
Two types of cleaning procedures have been applied to the data.
First the median statistics has been applied to remove about 5\% of data 
points as outlier individual occultation steps; these outliers are generally 
due to pulsar activity in individual detectors.
Second, contaminating  objects  in the limbs to the source
being studied have been identified using a catalogue of flaring and 
bright objects detected at high energies. If present, these contaminating 
sources were included in the occultation step fit and 
consequently data corresponding to contaminating periods were removed.
In 7 cases (ESO 103-G35, IC 4329A, MCG-6-30-15, MKN 501,
NGC 4593, NGC 7172 and PKS 2155-304) contamination was found and the above
procedure applied. 
Of course any interfering sources that are not taken into 
account (i.e. are not included in the database) will systematically increase
the flux estimate. This maybe  particularly important for sources 
closer than 5$^{\circ}$ to the galactic plane or within a cone of 
45$^{\circ}$ around the 
galactic center up to 30$^{\circ}$ (Connaughton et al. 1998a).
Since we are probing the extragalcatic sky, only a handfull of sources 
(ESO103-G35, ESO141-G55, Fairall 49 and MKN509) fall within the above cone and
in anycase, all of them are located just at the extreme boundaries of the 
cone and therefore they have not been excluded from the present analysis. 
Moreover, comparison of our fluxes with OSSE and BeppoSAX 
data for two of the above sources suggests that contamination from galactic
objects  was in these cases negligible.
\\
Fluxes in the 20-100 keV energy band have been obtained by
folding a single power law of photon index 1.75 for Seyfert galaxies, 3C273 
and M82, and  2.25 for 
BL Lac objects, with the BATSE instrumental response function
and then computing weighted mean flux values over the whole observation period.
Although these mean estimates depend on the choice of the photon index, the 
differences in fluxes are of the order of 10-20$\%$  around the 
values reported in Table 1 for $\Delta$ $\Gamma$= $\pm0.25$.

Before proceedings with the discussion of our results and in view of the
difficulties associated with the data analyisis, it is important to 
search for possible systematic effects on the flux measurements.
Two types of systematic
errors are particularly relevant: those affecting the overall 
normalization (which are particularly important for the comparison 
with other instruments), and those affecting the size of the 
fluctuations (which are relevant to estimate the confidence level 
of a detection).
Comparison of BATSE occultation results with contemporaneous observations 
with other CGRO instruments have indicated that the BATSE flux values
maybe systematically higher. In the most extensive study published to date, 
Much
et al. (1996) found that the BATSE Crab Nebula flux was $\sim$20\% higher
than OSSE flux in the same energy range. More recently, Parsons et al. (1998)
compared BATSE measurements of NGC 4151 with those of OSSE and derived a 
BATSE/OSSE flux ratio of 1.45$\pm$0.08. In view of the relevance 
of the Parsons et al. study to the present one, we have performed a similar
analysis on all the sample sources for which contemporaneous BATSE/OSSE
data (kindly provided to us by the OSSE team, Johnson 1997, private 
comunication) could be analyzed . The results of this analysis are
summarized in the parameter R reported in Table 1. 
For every contemporaneous BATSE/OSSE observation we have estimated the 
relative flux ratio in the 20-100 keV band and than we have calculated 
R which represents the weighted mean 
value with its associated error of all the available flux ratios.
The numbers
in parentheses indicate how many periods were examined (first quote) and the 
total number of observation days analyzed over those periods (second quote).
It is evident from Table 1 that R is  $\ge$ 1 in most cases, the
weighted mean value of R being 1.36 $\pm$ 0.03 (this value becomes 
1.5 $\pm$ 0.13 if the two strongest sources, NGC4151 and 3C273, 
are removed). Thus our result confirms previous reports of a BATSE/OSSE
normalization discrepancy (Much et al. 1996, Parsons et al. 1998) and
further indicates that in the particular case of extragalactic studies
the BATSE flux can be sistematically higher that the OSSE one by as much as 
35$\%$. The possible origin for this normalization problem is at the moment 
under investigation and will hopefully be solved in the near future.
A possible  explanation maybe source interference which is not taken into 
account in the data cleaning. This is an important
problem to address if BATSE light curves are to be used to monitor
bright source behaviour at high energies. 
\\
We have also compared BATSE data to published BeppoSAX observations,
although in this case the comparison is hampered by the lack of simultaneous
data. Nevertheless we get a BATSE/SAX flux ratio in the range 1.6-1.8 
which again
indicates that our fluxes maybe  overestimating the source high
energy emission.  However, since for individual sources the  
discrepancies found can be less
than quoted and in fact $\sim$ 1 (see for example 3C273 and other sources
in Table 1), no correction is applied 
to the BATSE data in the present work. Furthermore, as the errors too
are effected in the same way as the fluxes, this normalization problem has no 
impact on the detection level of a given source.  
\\
The effects of systematic errors have also been carefully considered by 
examining "blank fields" randomly distributed around the sky.
It was found that while the flux values for each field were consistent
with the quoted errors, the spread around zero of the mean fluxes from a large
number of these empty fields was about 80$\%$ higher than would be
expected by the statistical errors alone (Stephen 1998, private comunication).
For this reason all the errors 
on the mean fluxes were adjusted by 80$\%$ so as to be conservative in the 
source detection estimate. This estimate of the systematic errors is slightly 
higher than the 65$\%$ obtained by Connaughton et al. (1998a), but compatible
given the difference in the number of fields analysed.
\\
Table 1 lists all sources contained in the sample together with their optical
classification, X-ray data in the 2-10 keV band, a compendium of the 
results of the BATSE data analysis (flux and significance of
detection) and the BATSE/OSSE flux ratio.
Significant detection at $\ge$ 10$\sigma$ level has been found for three
sources (NGC 4151, 3C 273 and NGC5506), while at the 
5$\sigma$ level the number of detected sources grows to 14. Marginal
detections (3$<\sigma<$5) can be claimed for 13  sources while only 9
galaxies were observed below the 3 $\sigma$ confidence level.

\section{Discussion}

The main result of this work is the detection of high-energy emission
from the majority of the sample sources.
Comparison between OSSE and BATSE data indicates that 6 objects are here
reported as hard X-ray emitting sources for the first time:
ESO198-G24, H0917-074, ESO 103-G35, H1846-786,
3C 445 and  MKN1152, all detected above the 3 $\sigma$ confidence
level; of these only ESO 103-G35 has been reported as a high energy source
by BeppoSAX (Antonelli et al. 1998).
It is also interesting to find out that 5 out of 8 Seyferts 2 in the
sample have been detected, in agreement with the expectation of the
unified theory, which predicts similar energy output for 
type 1 and type 2 Seyferts at high energies. 
Particularly interesting is also the BATSE detection of
two BL Lac objects out of four in the sample: MKN 501 
and PKS 2155-304. Due to the steepness of their spectra compared to Seyferts,
BL Lacs have less probability of being detected by BATSE unless higher
than expected brightness was present during BATSE monitoring (see below).
\\
In order to assess the reliability of our results and 
to investigate the approximate spectral shape of AGN in the 
BATSE energy band, a comparison with soft 
X-ray  fluxes  has been performed.
BATSE flux estimates, however, are the result of a long
integration period (years) and thus represent averaged values over this
period, while observations with pointed instruments provide
instantaneous flux measurements. 
This has been taken into account by using in the 2$-$10 keV band  
mean flux values as estimated from data in the  literature 
(Ciliegi et al. 1993,
Malaguti et al. 1994, Malizia et al. 1997, Polletta et al. 1996).
The errors on this mean flux has been set  equal to 
the observed range of X-ray variability or  VR.
\\
In Figure 1, BATSE flux values are plotted against these
2-10 keV flux measurements. Lines plotted in the figure correspond to
the ratio expected for two values of the photon index (1.5 and 2) under the 
hypothesis of a single power law continuum from 2 to 100 keV.
It is interesting to find that a good fraction of galaxies 
fall within the region
constrained by these two spectral indices,  particularly if the range 
of X-ray flux variability is taken into account. More specifically the 
broad band  (2-100 keV) power law slope can be estimated from the fluxes
emitted in the two bands: while the average photon index is 1.7, the 
weighted mean value is 1.6$\pm$0.1 
(considering only detection above 3$\sigma$ and excluding BL Lacs); 
removal of NGC4151 and 3C273 from the sample does not change this mean. 
Decreasing the BATSE fluxes by the amount required by the
BATSE/OSSE comparison would make this mean photon index slightly steeper. 
Therefore  the canonical 1.7 power law  which best represents the X-ray data
gives  also a good description of the spectrum in the 20-100 keV band.
This result indicates
that the reflection component which is responsible for flattening the 
intrinsic spectrum in the 2-10 keV band, is relevant also above 20 keV and 
present in most sample sources. Furthermore, BATSE observations imply that
spectral breaks and/or
steepening  must occur preferentially above about 100 keV for consistency 
with our findings. This is in line with BeppoSAX results on Seyfert 
galaxies which locate the spectral cutoff typically  at energies
$\ge$ 100 keV (Matt 1998, Bassani et al. 1998).
\\
A couple of exceptions are however worthy of a note. Two sources, 
MCG-6-30-15 and NGC2992, are located in the region characterized 
by spectral indices greater than 2.0 (or, alternatively, in these sources the
BATSE flux is underestimated, which is a bit surprising in view of the above 
discussion). NGC2992 is however compatible with
flatter indices, if one considers the source's large range in variability 
and, in particular, the gradual decline
in flux by a factor of $\sim$ 20 monitored  over the last 10 years 
(Polletta et al. 1996):
by sampling the last part of this period BATSE probably underestimates the
source average flux. The case of MCG-6-30-15 is less obvious and deserves 
further investigation: either the source was dimmer during the BATSE coverage
with respect to previous 2-10 keV observations or a break below 100 keV
(Molendi et al. 1998) must be postulated. 
Since BeppoSAX observation of this source locates the
break at energies $\ge$ 100 keV (Molendi et al. 1998),  the first 
alternative is more
plausible suggesting  that  BATSE can indeed be used as a monitor 
of bright source states.
The location of MKN501 and 2A 1219+305 is also peculiar as these two objects
have unusually flat spectra for BL Lac
objects (Ciliegi et al. 1995). While the case of 2A 1219+305 is less stringent,
the detection being only an upper limit, the case of MKN501 is particularly
interesting. Strong flaring activity has been reported from this source at
high energies during the BATSE monitoring period (Pian et al. 1998 and Catanese
et al. 1997, Lamer et al. 1998); analysis of the BATSE light curve 
indicates indeed that flare like events are present along with a gradual 
increase in flux during April-July 1997
(Connaughton et al. 1998a,b). During at least one such event (in April) 
the source spectrum was
flatter than previoulsy reported (Pian et al. 1998). Given the peculiar
behaviour observed during BATSE coverage, a proper comparison
between low and high energy bands  should use contemporaneous data.
\\
Finally, BATSE results have been compared with previous
studies of the average AGN flux at high energies; the average flux of the 
Seyfert galaxy population as a whole has
been found to be  (1.74$\pm$0.05) $\times$ 10$^{-5}$ photons/cm$^{2}$ s keV 
in the 20-100 keV band and if strong sources such as
NGC 4151, 3C 273 and IC 4329A are excluded, the mean flux reduces to
(1.24$\pm$0.05) $\times$ 10$^{-5}$ photons/cm$^{2}$ s keV. 
Dividing the sample in classes gives a mean flux of (1.88$\pm$0.05) $\times$ 
10$^{-5}$ photons/cm$^{2}$ s keV for Seyfert 1, (1.31$\pm$0.11) $\times$ 
10$^{-5}$ photons/cm$^{2}$ s keV for Seyfert 2  and (1.03$\pm$0.19) 
$\times$ 10$^{-5}$ photons/cm$^{2}$ s keV for BL Lac
objects, in agreement with the results of Maisack, Wood and
Gruber (1994) based on HEAO A4 data, albeit a 40$\%$ reduction in the BATSE
fluxes  would improve the comparison. 
Also, if only the radio
quiet Seyfert 1 galaxies in the sample (12 objects) are
taken into account, the average flux is (1.22$\pm$0.08) $\times$ 10$^{-5}$ 
photons/cm$^{2}$ s keV, compatible within errors with 
the value obtained by Gondek et
al. (1996) for the same class of objects; in this case no correction to a 
lower flux level is required by the comparison.
While the agreement found gives 
confidence to our data analysis (the level of each detection is a 
sound result, while the flux estimate maybe in some cases slightly 
overestimated) it also demonstrates the 
potential of BATSE for cumulative studies of selected samples of objects
for example in search of a population of objects bright at high energies.
Down to a limiting flux  of 7.8 10$^{-11}$ erg cm$^{-2}$ s$^{-1}$ BATSE 
has been able to detect 14 objects above 5 $\sigma$ level over 8.2 sr.
This is to be considered a lower limit to the number of high energy
emitting AGN as highly absorbed 
objects, i.e. objects missed in the Piccinotti survey for their absorption
in excess of  10$^{24}$ cm$^{-2}$ but visible above 10 keV (Bassani, Cappi, 
Malaguti 1998), are probably not included. This suggests that an 
improvement in sensitivity of a factor of $\sim$10 as foreseen for
the next generation of high energy telescopes such as INTEGRAL (Winkler 1996)
would allow more than  600 objects to  be visible over the entire sky
at  $\ge$ 5 $\sigma$ level.

{\bf Acknowledgement}

We thank N. Johnson for use of OSSE data prior to publication.\\
It is a pleasure to thank M. Malaspina for his support in the implementation
of the BATSE analysis package at TeSRE and J.B.Stephen for his work
on the evaluation of systematic errors in BATSE data analysis.\\
A.M.  acknowledges the CNR and Southampton University for a fellowship.
Financial support by MURST, CNR and ASI is gratefully acknowledged.
\newpage

\begin{deluxetable}{lllrlrrr}
\tablecolumns{8}
\tablewidth{0pc}
\tablecaption{}
\tablehead{
\colhead{}  &\colhead{}  &  \multicolumn{2}{c}{Low Energy}   & 
\multicolumn{4}{c}{High Energy} \\
\cline{3-4}  \cline{6-8} \\
\colhead{Source} & \colhead{Type} &\colhead{F$_{MEAN}^{\dagger}$}
&\colhead{VR$^{\ddagger}$} & \hspace*{1cm} &
\colhead{F$_{BATSE}^{\dagger}$} &
\colhead{N$_{\sigma}^{BATSE}$} & \colhead{R$^{\star}$}  \nl  
\colhead{} &\colhead{} &\colhead{(2-10)keV} & \colhead{} & &
\colhead{(20-100)keV} & \colhead{} & \colhead{}   } 
\startdata
III ZW 2          & Sey 1   & \phn\phn2.93 & 0.64 &   &\phn4.88$\pm$1.62 
&\phn\phn3.02 & 1.80$\pm$1.0 (1-13) \nl 
MKN 1152          & Sey 1.5 & \phn\phn2.02 & 1.22 &   &\phn5.99$\pm$1.51 
&\phn\phn3.96 & \nodata \nl 
FAIRALL 9         & Sey 1   & \phn\phn2.89 & 2.15 &   &\phn3.80$\pm$1.70 
&\phn\phn2.24 & 1.65$\pm$1.37 (1-13) \nl 
NGC 526A          & Sey 2   & \phn\phn3.13 & 2.43 &   &\phn12.44$\pm$1.54  
&\phn8.07    & 4.40$\pm$1.5 (1-13) \nl 
MKN 590           & Sey 1.2 & \phn\phn1.97 & 0.94 &   &\phn6.75$\pm$1.47  
&\phn\phn4.59 & 4.67$\pm$4.8 (1-14) \nl 
ESO 198$-$G24     & Sey 1   & \phn\phn3.30 & 1.31 &   &\phn5.27$\pm$1.66  
&\phn\phn3.18 & \nodata \nl 
3C 120            & Sey 1   & \phn\phn3.94 & 1.26 &   &\phn4.84$\pm$1.58  
&\phn\phn3.07 & 3.41$\pm$2.5 (2-18)\nl 
PKS 0548$-$322    & BL Lac  & \phn\phn2.73 & 1.17 &   & $<$3.76         
&\phn\nodata & \nodata \nl 
H0557$-$385       & Sey 1   & \phn\phn3.17 & 1.75 &   &\phn4.74$\pm$1.53   
&\phn3.09  & \nodata \nl 
H0917$-$074       & Sey 1   & \phn\phn1.88 & 1.08 &   &\phn5.63$\pm$1.55  
&\phn\phn3.63 & \nodata  \nl 
NGC 2992          & Sey 2   & \phn\phn5.03 & 4.58 &   & $<$3.18
&\nodata & \nodata \nl 
M 82              & SB      & \phn\phn2.52 & 0.16 &   &\phn5.94$\pm$1.12  
&\phn\phn3.08 & \nodata \nl 
NGC 3227          & Sey 1.5 & \phn\phn3.15 & 2.35 &   &    11.26$\pm$1.53
&\phn7.36    & 1.67$\pm$0.6 (4-51)\nl 
NGC 3783          & Sey 1   & \phn\phn4.69 & 3.35 &   &    13.00$\pm$1.64 
&\phn7.91    & \nodata \nl 
NGC 4151          & Sey 1.5 & \phn22.55    & 15.05&   &    77.46$\pm$1.59 
&\phn48.77& 1.37$\pm$0.03 (6-94)\nl 
2A 1219+305       & BL Lac  & \phn\phn2.98 & 1.38 &   &\phn5.70$\pm$2.29  
&\phn\phn2.48 & \nodata \nl 
3C 273            & QSO     & \phn11.68    & 6.14 &   &    39.13$\pm$1.59
&\phn24.58    & 1.04$\pm$0.1 (5-43)  \nl 
NGC 4593          & Sey 1   & \phn\phn3.17 & 1.17 &   &\phn9.81$\pm$1.59  
&\phn\phn6.16 & \nodata \nl 
MCG$-$6$-$30$-$15 & Sey 1   & \phn\phn5.77 & 1.51 &   & $<$4.30
&\nodata   & 1.08$\pm$1.6 (1-7)\nl 
IC 4329A          & Sey 1   & \phn12.15    & 7.05 &   &    19.44$\pm$1.99
&\phn9.78    & 1.34$\pm$1.19  (6-71)\nl 
NGC 5506          & Sey 1.9 & \phn\phn6.52 & 4.75 &   &    19.16$\pm$1.60
&\phn11.97    & 2.57$\pm$0.56  (3-29)\nl 
&             &             &              &      &   &     & \nl
NGC 5548          & Sey 1   & \phn\phn4.35 & 2.37 &   &\phn10.53$\pm$1.58 
&\phn6.67    & 1.45$\pm$0.63  (3-26)\nl 
MKN 501           & BL Lac  & \phn\phn3.40 & 1.25 &   & 10.31$\pm$2.77
&\phn3.71   & \nodata \nl 
FAIRALL 49        & Sey 2   & \phn\phn1.94 & 1.43 &   & \phn3.98$\pm$1.86    
&\phn2.14 & \nodata \nl 
ESO 103$-$G35     & Sey 2   & \phn\phn2.09 & 0.89 &   &\phn7.92$\pm$1.86  
&\phn\phn4.26 & \nodata  \nl 
H1846$-$786       & Sey 1   & \phn\phn2.46 & 0.76 &   &\phn5.99$\pm$1.76
&\phn\phn3.39 & \nodata \nl 
ESO 141$-$G55     & Sey 1   & \phn\phn3.26 & 0.44 &   & $<$3.50 
&\nodata & \nodata \nl 
MKN 509           & Sey 1   & \phn\phn4.41 & 1.29 &   &    12.47$\pm$1.52 
&\phn8.18    & 0.96$\pm$0.6 (2-20)\nl 
PKS 2155$-$304    & BL Lac  & \phn\phn9.29 & 5.21 &   &\phn7.70$\pm$1.94
&\phn\phn3.97 & 1.19$\pm$0.82 (2-14) \nl 
NGC 7172          & Sey 2   & \phn\phn3.06 & 1.29 &   &\phn8.00$\pm$1.55 
&\phn5.15    & 1.16$\pm$0.37 (3-26) \nl 
NGC 7213          & Sey 1   & \phn\phn3.66 & 1.34 &   &\phn3.39$\pm$1.65 
&\phn\phn2.05 & 2.23$\pm$1.5 (1-7) \nl 
3C 445            & Sey 1   & \phn\phn2.28 & 1.09 &   &\phn12.15$\pm$1.54 
&\phn7.88    & \nodata \nl 
NGC 7314          & Sey 1.9 & \phn\phn2.76 & 1.46 &   &\phn4.19$\pm$1.58  
&\phn\phn2.66 & \nodata  \nl 
NGC 7469          & Sey 1   & \phn\phn3.02 & 1.22 &   &\phn7.77$\pm$1.51 
&\phn\phn5.15 & 2.39$\pm$2.05 (1-7) \nl 
MCG$-$2$-$58$-$22 & Sey 1.5 & \phn\phn4.08 & 2.32 &   &\phn7.53$\pm$1.53  
&\phn\phn4.93 & \nodata  \nl 
NGC 7582          & Sey 2   & \phn\phn3.60 & 3.00 &   &\phn8.97$\pm$1.65  
&\phn\phn5.43 & $<$2.14 (1-7)  \nl 
\enddata

\footnotesize
\tablenotetext{(\dagger)}{ = flux in units of 10$^{-11}$ erg cm$^{-2}$ 
s$^{-1}$.\\ 
BATSE fluxes are the weighted mean calculated over nearly four 
year of observatios} 
\tablenotetext{(\ddagger)}{ = fluxes variability range in the 2-10 keV 
energy band} 
\tablenotetext{(\star)}{ = weighted mean value of the BATSE/OSSE flux 
ratios in the 20-100 keV with its associated error for contemporaneous 
observations. The numbers in parentheses indicate how many periods were 
examined (first quote) and the total number of observation days analyzed over 
those periods (second quote).}
\tablenotetext{\star\star}{ upper limits are at 2$\sigma$ level} 

\end{deluxetable}

\begin{figure}
\plotone{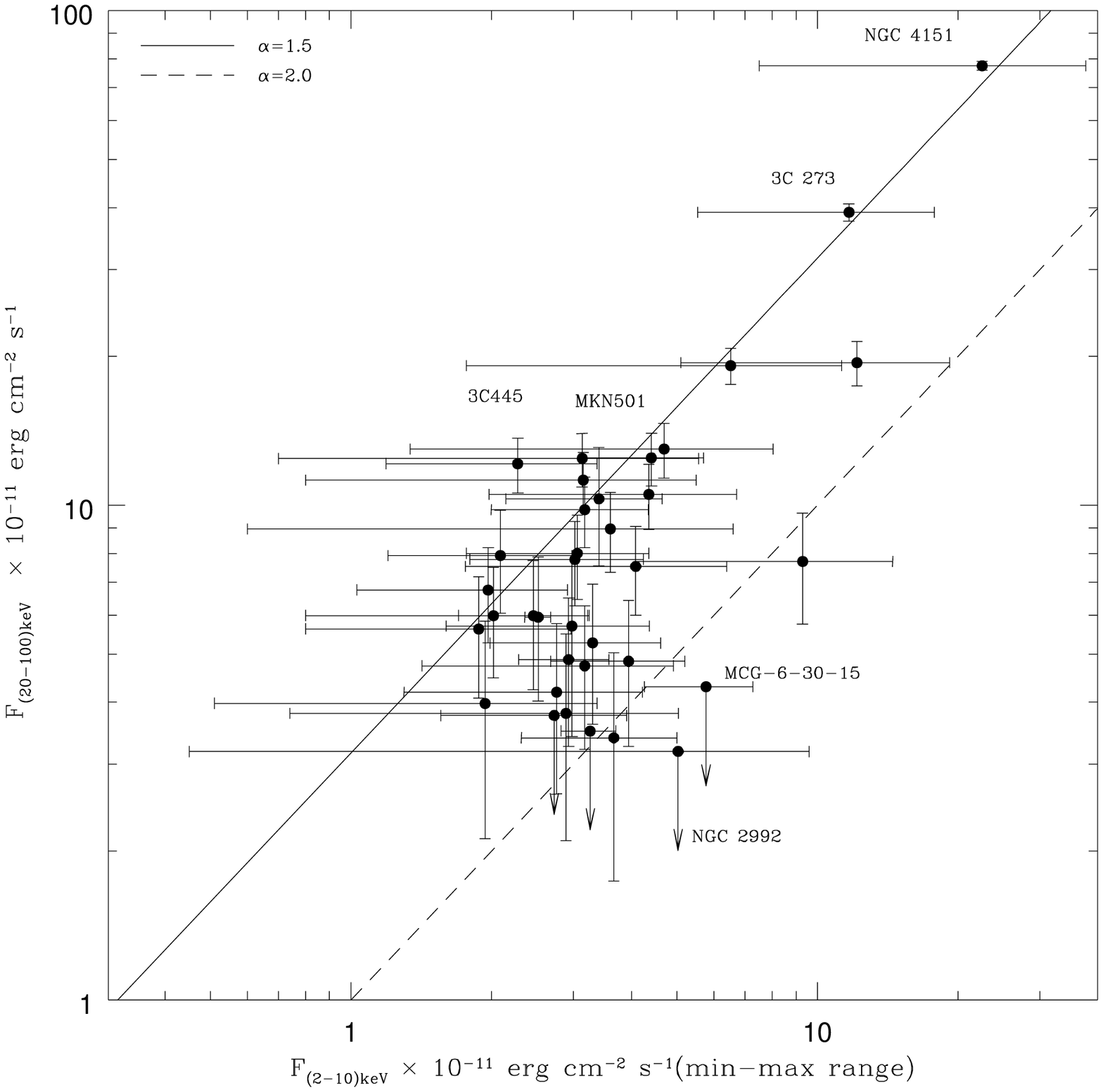}
\caption{BATSE 20-100 keV weighted mean fluxes on a $\sim$ 3 years of
observations with their relative errors, are plotted against 2-10 keV 
average fluxes.
The horizontal bars in the figure are the 2-10 keV flux ranges
(minimum and maximum value found in the literature). 
Lines plotted correspond to the ratio between fluxes expected for
various values of photon index under the hypothesis of a single power
law continuum from 2 to 100 keV. 
Although 3C445 has a slop flatter than 1.5 this is consistent with recent X-ray
measurements (Sambruna et al. 98).}
\end{figure}

\end{document}